\documentclass[letterpaper, 10 pt, conference]{ieeeconf}  
\IEEEoverridecommandlockouts                              
\usepackage{graphicx}
\usepackage{epstopdf}
\usepackage{amsfonts}
\usepackage{amsmath,amssymb}
\usepackage{color}
\usepackage{tikz}
\usetikzlibrary{automata}
\usepackage{color}
\usepackage{verbatim}
\usepackage{comment}
\usepackage{algorithm}
\newtheorem{definition}{Definition}

\newtheorem{lemma}{Lemma}

\newtheorem{proposition}{Proposition}
\newtheorem{theorem}{Theorem}

\newtheorem{problem}{Problem}

\usepackage{varioref}

\begin{document}
\title{Timed Supervisory Control for Operational Planning and Scheduling under Multiple Job Deadlines}
\author{Ahmad Reza Shehabinia, Liyong Lin, Rong Su
\thanks{Ahmad Reza Shehabinia, Liyong Lin, Rong Su are affiliated with School of Electrical and Electronic Engineering, Nanyang Technological University, 50 Nanyang Avenue, Singapore 639798. Emails: \{ahmadrez1,llin5,rsu\}@ntu.edu.sg.}\\
}

\maketitle
\begin{abstract}
In this paper we model an operational planning and scheduling problem under multiple job deadlines in a time-weighted automaton framework. We first present a method to determine whether all given job specifications and deadlines can be met by computing a supremal controllable job satisfaction sublanguage. When this supremal sublanguage is not empty, we compute one of its controllable sublanguages that ensures the minimum total job earliness by adding proper delays. When this supremal sublangauge is empty, we will determine the minimal sets of job deadlines that need to be relaxed.
\end{abstract}

{\em Index Terms} -- time-weighted automaton, controllability, scheduling, earliness, timed supervisory control

\section{Introduction\label{GD}}
Job deadline satisfaction problems have been extensively studied in the operations research community due to their important applications in manufacturing and logistics. There is an enormous number of publications on this subject. For example, \cite{Pin08} provides a thorough account on job shop scheduling problems with deadlines. \cite{SS07} gives a survey of scheduling with controllable processing times, in order to meet deadlines in an optimal manner, e.g., reducing job earliness. When some deadlines are deemed infeasible to be fulfilled, relaxations can be considered, usually in an optimal manner. For example, in \cite{HRM05} the authors consider a single machine scheduling problem with common due dates, and the performance is measured by the minimization of the sum of earliness and tardiness penalties of the jobs. 
In operations research usually mathematical and constraint programming models are used.

Recently, researchers in the supervisory control community have been formulating the job deadline satisfaction problems in automaton (or language) based models, motivated by the belief that supervisory control theory \cite{RW87} \cite{WR87} can ensure optimal performance while guaranteeing operational safety and liveness. In \cite{cury2013scheduling} the authors adopt the Brandin-Wonham timed control paradigm~\cite{BW94} to deal with supervisory control for job deadlines in cluster tools applications, whose modeling formalism follows the Ostroff's semantics for timed transition models~\cite{Ost89}. In~\cite{BBHM05} the authors adopt the timed automaton modeling formalism~\cite{AD94} and use reachability analysis techniques to determine deadline satisfaction in an industrial case study. In \cite{SSR11} a new time optimal supervisory control paradigm is introduced, in which a plant is modeled as a time-weighted automaton and a logic requirement is modeled as an unweighted automaton. The heap-of-pieces theory \cite{Vie97} is used to describe execution time of each trajectory in a quasi-concurrent setup, i.e., executions of different transitions in the trajectory can be overlapped but their starting moments must be sequentially ordered. The goal in \cite{SSR11} is to achieve minimum makespan.

In this paper we adopt the time-weighted automaton modeling formalism introduced in \cite{SSR11} to address the job deadline satisfaction problems. The basic setup consists of a plant modeled as a time-weighted automaton, a logic requirement language that takes care of general logic constraints about operational safety and progress, and a collection of job requirements associated with relevant job deadlines. We first tackle the problem of determining the least restrictive controllable sublanguage (denoted as $S$ for easy reference) of the plant that satisfies all requirements and deadlines. 
When $S$ exists, our second problem is to determine an optimal strategy to add delays to relevant transitions of the plant so that there exists the least restrictive controllable sublanguage that satisfies all requirements and deadlines with minimum job earliness. By solving a constraint optimization problem we show that the set of all optimal delays can be computed. 
When $S$ does not exist due to infeasibility of fulfilling some deadlines, our next problem is to determine how to relax some deadlines and in order to minimize the impact of deadline relaxation, we present an algorithm to determine the minimal sets of deadlines which need to be relaxed.


We organize the paper as follows. Firstly, we review some basic  concepts of time-weighted automaton formalism introduced in \cite{SSR11} in Section II. Then we present the baseline problem of synthesizing the supremal controllable job-satisfaction sublanguages in Section III. After that, we bring in the problem of delay addition for minimum job earliness in Section IV, and the problem of relaxing job deadlines in Section V. After providing a simple illustration example in Section VI, conclusions are drawn in Section VII.

\section{Time-weighted automaton}
We assume that the reader is familiar with concepts and operations in supervisory control theory \cite{RW87} and follow the notations in \cite{Won13}. Given two strings $s,t\in\Sigma^*$, we write $s\leq t$ to denote
$s$ being a \emph{prefix substring} of $t$. Given a language $L\subseteq\Sigma^*$ we use $\overline{L}$ to denote its \emph{prefix closure}. Given two
languages $L,L'\subseteq\Sigma^*$, let $LL'$ denote {{ } their concatenation}. Given an arbitrary set $S$ we use $|S|$ to denote its cardinality. We write
$P:\Sigma^*\rightarrow\Sigma'^*$ for the \emph{natural
projection} with respect to $(\Sigma,\Sigma')$, where $\Sigma'\subseteq\Sigma$.  {{ }The projection is naturally extended to mapping between languages. The inverse of $P$ is denoted by $P^{-1}$}. Given
$L_1\subseteq\Sigma_1^*$ and $L_2\subseteq\Sigma_2^*$, we write {{ } $L_1||L_2:=P_1^{-1}(L_1) \cap P_2^{-1}(L_2)$} for {{ } their} \emph{synchronous product}.
Let $\mathbb{R}$, $\mathbb{R}^+$ and $\mathbb{N}$ {{ } denote} the collections of reals, positive reals and natural numbers respectively. A \emph{finite-state time-weighted automaton} is a 3-tuple
$(G=(X,\Sigma,\xi,x_0,X_m),f,h)$, {{ } where $G$ is a finite-state automaton; $f:\mathcal{T} \rightarrow\mathbb{R}^+$ is the weight function, where $\mathcal{T}:=\{(x,\sigma)\in X\times\Sigma \mid \xi(x,\sigma)!\}$ denotes the set of transitions defined in $G$, that assigns to each transition of $G$ a finite positive real representing the duration required for the corresponding transition to be completed; $h \subseteq \Sigma\times\Sigma$ is the reflexive and symmetric mutual exclusion relation. A pair $(\sigma,\sigma')\in h$
 if and only if $\sigma, \sigma'$ cannot be under execution simultaneously.
For notational simplicity, we write $(\sigma,\sigma')\in h$ to denote both
$(\sigma,\sigma')\in h$ and $(\sigma',\sigma)\in h$. Let
$L(G):=\{s\in\Sigma^* \mid \xi(x_0,s)!\}$ be the \emph{closed} behavior
of $G$ and $L_m(G):=\{s\in L(G) \mid \xi(x_0,s)\in X_m\}$ be the
\emph{marked} behavior of $G$. Let $\phi(\Sigma)$ and $\varphi(\Sigma)$ denote respectively the set of
finite-state automata, and the set of finite-state time-weighted automata, whose alphabets are $\Sigma$}. 
We adopt the time-stamp interpretation of string execution time and the approach based on the heaps-of-pieces theory to calculate the string execution time $\upsilon_{G,f,h}(s)$~\cite{SSR11}.

Given $(G=(X,\Sigma,\xi,x_0,X_m),f,h)$, we use $\varpi:L(G)\rightarrow (X\times\Sigma)^*$ to map each string in $L(G)$ to the (unique) trajectory (or path) in $G$, where $\varpi(\epsilon):=\epsilon$. By \cite{SSR11} we know that the mutual exclusion relation $h$ induces a resource set $\mathcal{R}$ and a map $R:\mathcal{T}\rightarrow 2^{\mathcal{R}}$, which maps each transition $\tau\in\mathcal{T}$ to its set of resources. Let $n_h=|\mathcal{R}|$. Then there exists a morphism $\hat{\mathcal{M}}_{G,f,h}:\mathcal{T}^*\rightarrow\mathbb{R}_{max}^{n_h,n_h}$, where $\mathbb{R}_{max}^{n_h,n_h}$ is the collection of all matrices whose dimensions are $n_h\times n_h$ such that $\hat{\mathcal{M}}_{G,f,h}(\epsilon)$ is defined as the unit matrix $\textbf{I}_{n_h\times n_h}$, i.e., all diagonal entries are 0 and all other entries are $-\infty$, and for each $\tau\in\mathcal{T}$,
\begin{center}$\hat{\mathcal{M}}_{G,f,h}(\tau)_{qv}:=\left\{\begin{array}{ll} 0 & \textrm{if $q=v$, $q\notin R(\tau)$, or $q\in R(\tau)$, $v\notin R(\tau)$}\\
f(\tau) & \textrm{if $q,v\in R(\tau)$}\\
-\infty & \textrm{otherwise}\end{array}\right.$\end{center}
In \cite{SSR11} it has been shown that
\begin{equation}\upsilon_{G,f,h}(s)=\textbf{1}_{n_h}^t\hat{\mathcal{M}}_{G,f,h}(\varpi(s))\textbf{1}_{n_h},\end{equation}
where $\textbf{1}_{n_h}\in\mathbb{R}_{max}^{n_h}$
is the $n_h$-dimensional column vector, whose
entries are all $0$, and $\textbf{1}_{n_h}^t$ is the transpose of
$\textbf{1}_{n_h}$. When the context is clear, we use $\upsilon_f(s)$ to denote $\upsilon_{G,f,h}(s)$. 

Let $\Sigma=\Sigma_c\dot{\bigcup}\Sigma_{uc}$, where
$\Sigma_c$ and $\Sigma_{uc}$ denote respectively the sets of
\emph{controllable} events and \emph{uncontrollable}
events.

\begin{definition}\label{Def10}\textnormal{\cite{RW87} Given $G\in \phi(\Sigma)$, $K\subseteq L_m(G)$ is \emph{controllable} with respect to $G$ if
$\overline{K}\Sigma_{uc}\cap L(G)\subseteq \overline{K}$.
\hfill $\Box$}
\end{definition}

\section{Supremal controllable job satisfaction sublanguage}
In this section, we propose an algorithm to compute the supremal controllable sublanguage that satisfies the job requirements and deadlines. To that end, we formally introduce the notion of job requirements and their deadlines below.

Let $\mathcal{E}_T=\{(E_i, d_i)\in 2^{{\Sigma_i^*}} \times \mathbb{R}^{+} \mid i\in I\}$, where $E_i$ is a regular language, called \textit{job requirement}, over $\Sigma_i \subseteq\Sigma$ and $d_i\in \mathbb{R}^{+}$ is its \emph{deadline} (or \emph{due date}). Here $I$ is a finite index set and each tuple $(E_i,d_i)\in \mathcal{E}_T$ is interpreted as follows: each job requirement $E_i$ has to be satisfied within its pre-specified due date $d_i$. In this work, the completion of any {\em process} $s \in E_i$ is said to satisfy or complete the job requirement $E_i$. Since an operational sequence generated by the plant may involve events that contribute to the satisfaction of different jobs, 
we also say an operation sequence $s\in \Sigma^*$ completes job $E_i$ if $P_i(s)\in E_i$, where $P_i:\Sigma^*\rightarrow \Sigma_i^*$ is the natural projection {{ } with respect to $(\Sigma, \Sigma_i)$}. We define the {job $E_i$ execution time by $s$ (under $f$)} as \[t_{E_i, f}(s)=\min_{s'\leq s\wedge P_i(s')=P_i(s)}\upsilon_f(s')=\max_{s'\leq s\wedge s'\in \Sigma^*\Sigma_i}\upsilon_f(s'),\]
i.e., the job $E_i$ execution time by $s$ is the duration from the beginning of $s$ to the last event of $s$ in $\Sigma_i$. For example, if $\Sigma=\{a,b,c,d\}$, $\Sigma_i=\{a,c\}$, $s=abcd$, then $t_{E_i, f}(s)=\upsilon_f(abc)$ is the job $E_i$ execution time by $s$. Moreover, if $s$ completes $E_i$, i.e., $P_i(s) \in E_i$, then $t_{E_i, f}(s)$ is said to be the \emph{job $E_i$ completion time by $s$ (under $f$)}. {{ } For technical convenience, we assume $\Sigma=\bigcup_{i \in I}\Sigma_i$. We allow $\Sigma_i \cap \Sigma_j \neq \varnothing$ for $i \neq j$}. Hence each event executed by the plant contributes to the completion of some jobs and we allow the same event to contribute to the completion of several different jobs. 
Let
$W_f(\mathcal{E}_T):=\{s\in \Sigma^* \mid \forall i \in I, P_i(s) \in E_i \wedge t_{E_i, f}(s)\leq d_i\}$
be the (possibly empty) collection of all strings which satisfy all job requirements and the corresponding job deadlines. $W_f(\mathcal{E}_T)$ is a finite language since the transition duration of each event is positive and $\Sigma=\bigcup_{i \in I}\Sigma_i$. In addition to the job requirements, there is a \emph{general logic requirement} $E\subseteq\Sigma^*$ used to specify the safety properties and the progress properties. We now bring in the notion of a {\em controllable job satisfaction sublanguage}.

Given a time-weighted plant $(G,f,h)\in\varphi(\Sigma)$, the set of jobs and deadlines $\mathcal{E}_T=\{(E_i,d_i)\in 2^{\Sigma_i^*}\times\mathbb{R}^{+} \mid i\in I\}$ and a general logic requirement $E\subseteq\Sigma^*$, let
\begin{center}
$\mathcal{C}(G,f,h,E,\mathcal{E}_T):=
\{K\subseteq L_m(G)\cap E\cap W_f(\mathcal{E}_T)  \mid $\\ $K \textrm{ is controllable w.r.t. }G\}$
\end{center}
be the collection of all controllable sublanguages\footnote{We consider the framework of marking non-blocking supervisory control~\cite{Won13}. The work could be easily adapted to dealing with the non-marking non-blocking supervisory control framework.} of $L_m(G)$ satisfying all requirements and the job deadlines, i.e., the collection of {\em controllable job satisfaction sublanguages} of $(G,f,h)$ under $E$ and $\mathcal{E}_T$. There exists a unique element $K_*\in \mathcal{C}(G,f,h,E,\mathcal{E}_T)$ such that
\[\forall K\in \mathcal{C}(G,f,h,E, \mathcal{E}_T), K\subseteq K_*\]
We call $K_*$ the \emph{supremal controllable job satisfaction sublanguage of $(G,f,h)$ under $E$ and $\mathcal{E}_T$}, denoted as $\textrm{sup}\mathcal{C}(G,f,h,E,\mathcal{E}_T)$. The first problem is stated below.\\

\begin{problem}\label{prob1}\textnormal{Given a plant $(G,f,h)$, requirements $E$ and $\mathcal{E}_T$, compute $\textrm{sup}\mathcal{C}(G,f,h,E,\mathcal{E}_T)$.\hfill $\Box$}\\
\end{problem}

A solution to Problem~\ref{prob1} is given by the following algorithm. 

\begin{algorithm}
\caption{Compution of $\textrm{sup}\mathcal{C}(G,f,h,E,\mathcal{E}_T)$}
\label{algorithm1}
\begin{enumerate}
\item \textbf{Input}: A centralized model $(G,f,h)\in\varphi(\Sigma)$, requirements $E\subseteq \Sigma^*$ and $\mathcal{E}_T$ with $I=\{1, 2, \ldots, k\}$.
\item Construct a trim tree-structured automaton for $H=L_m(G) \cap E \cap W_f(\mathcal{E}_T)$ by constructing a tree-structured automaton for $L_m(G) \cap E \cap \lVert_{i \in I}E_i$ in breadth-first order with the following modifications:
\begin{enumerate}
\item Each state $q$ is maintained with a $(k+2)$-tuple $(\hat{\mathcal{M}}_{G,f,h}(s), \upsilon_f(s), t_1(q), t_2(q), \ldots, t_k(q))$, where $s$ is the unique string labeling the path from the root state $q_0$ to state $q$. Initially, $t_1(q_0)=t_2(q_0)=\ldots=t_k(q_0)=0$. 
\item For the immediate successor state $q'$ of $q$ via $\sigma$, let $\hat{\mathcal{M}}_{G,f,h}(s\sigma):=\hat{\mathcal{M}}_{G,f,h}(s)\hat{\mathcal{M}}_{G,f,h}(\sigma)$ and $\upsilon_f(s\sigma):=\textbf{1}_{n_h}^t\hat{\mathcal{M}}_{G,f,h}(s\sigma)\textbf{1}_{n_h}$. If $\sigma \in \Sigma_i$, let $t_i(q'):=\upsilon_f(s\sigma)$, else let $t_i(q'):=t_i(q)$. If $\exists i \in I, t_i(q) \geq d_i$, the subtree rooted at state $q$ is cut.
\item Repeat step b) for each reached but unexplored state $q$. 
\end{enumerate}
\item \textbf{Output}: The supremal sublanguage $K_*\subseteq H$ that is controllable w.r.t. $G$.
\end{enumerate}
\end{algorithm}

Here $t_i(q)$ is interpreted as the job $E_i$ execution time $t_{E_i, f}(s)$, where $s$ is the unique string labeling the path from the root state $q_0$ to state $q$. Since $H$ is a finite language, step 2) of the above algorithm clearly terminates. We immediately have the following result. \\
%


\begin{theorem}
Let $K_*$ be the output of Algorithm~\ref{algorithm1}. Then $K_*=\textrm{sup}\mathcal{C}(G,f,h,E,\mathcal{E}_T)$.\hfill $\Box$\\
\end{theorem}

{\em Proof}: The proof is straightforward. \hfill $\blacksquare$

\section{Supremal minimum-earliness controllable job satisfaction sublanguage}
In Problem \ref{prob1}, if $\textrm{sup}\mathcal{C}(G,f,h,E,\mathcal{E}_T)\neq\varnothing$, it is desirable to synthesize a non-empty controllable sublanguage with minimum job earliness, by adding proper delays to the occurrences of some controllable transitions. To formalize this idea we need to introduce the concept of \emph{earliness of job $E_i$ completion (under $f$)}.\\

\begin{definition}
\label{def: hard}
\textnormal{Given a string $s\in \Sigma^*$ and a job requirement $(E_i,d_i)$ such that $P_i(s) \in E_i$ and $t_{E_i, f}(s)\leq d_i$.
The \emph{earliness} of completing $E_i$ by $s$ (under $f$) is defined as $e_{i,f}(s)=d_i-t_{E_i, f}(s)$.
Accordingly, the \emph{earliness} of completing $\mathcal{E}_T$ by $s$ is defined as $e_f(s)=\sum_{i\in I} e_{i, f}(s)$, if $s \in W_f(\mathcal{E}_T)$.\hfill $\Box$}\\
\end{definition}


\begin{definition}\textnormal{
Given any finite language $K\subseteq W_f(\mathcal{E}_T)$, the \emph{earliness} of $K$ w.r.t. $\mathcal{E}_T$ is defined as $e_f(K):=\max_{s\in K} e_f(s)$. \hfill $\Box$}\\
\end{definition}

The earliness of the empty set is defined to be zero, i.e., $e_f(\varnothing):=0$. If $\textrm{sup}\mathcal{C}(G,f,h,E,\mathcal{E}_T)\neq\varnothing$, it is always possible to compute a non-empty controllable sublanguage with minimum job earliness by adding delays. To introduce delays to the plant model we have the following construction. {{ } For a given $(G,f,h) \in\varphi(\Sigma)$, let $D: \mathcal{T} \rightarrow \mathbb{R}^{+}\cup\{0\}$ be a weight function such that $D(x, \sigma)=0$ for $\sigma \in \Sigma_{uc}$ and $D(x, \sigma) \in \mathbb{R}^{+} \cup\{0\}$ for $\sigma \in \Sigma_{c}$}. Let ${\bf DC}$ be the collection of delays that satisfy above constraint. For each $D \in {\bf DC}$, the firing duration of each transition $(x, \sigma) \in \mathcal{T}$ is extended to $(f+D)(x,\sigma):=f(x,\sigma)+D(x,\sigma)$, where {{ }
$f+D:\mathcal{T} \rightarrow\mathbb{R}^{+}$ 
is an \emph{extended weight function} parameterized by $D\in {\bf DC}$}. Recall that, when there is no delay, to compute string execution time $\upsilon_{G,f,h}(s)$ we need to bring in an induced morphism $\hat{\mathcal{M}}_{G,f,h}$. After introducing the delay $D$, the induced morphism becomes $\hat{\mathcal{M}}_{G,f+D,h}$. Here for each $\tau\in\mathcal{T}$ we have
\begin{center}$(\hat{\mathcal{M}}_{G,f+D,h}(\tau))_{qv}:=\left\{\begin{array}{ll} 0 & \textrm{if $q=v$, $q\notin R(\tau)$, or $q\in R(\tau)$, $v\notin R(\tau)$}\\
(f+D)(\tau) & \textrm{if $q,v\in R(\tau)$}\\
-\infty & \textrm{otherwise}\end{array}\right.$\end{center}
which can be used in computing the new string execution time $\upsilon_{G,f+D,h}(s)$. The computation of earliness is also affected accordingly and changed from $e_f$ to $e_{f+D}$ after adding the delay. The existence of the \emph{supremal minimum-earliness
controllable job satisfaction sublanguage} is ensured by the following proposition. \\


\noindent \begin{proposition} \label{supre} \textnormal{Let $D \in {\bf DC}$ be a delay function. If $\textrm{sup}\mathcal{C}(G,f+D,h,E,\mathcal{E}_T)\neq\varnothing$, then there
exists a non-empty set $\hat{K}_*\in\mathcal{C}(G,f+D,h,E,\mathcal{E}_T)$ such that
for any non-empty $K\in\mathcal{C}(G,f+D,h,E,\mathcal{E}_T)$ the
following holds,\\
\begin{enumerate}
\item $e_{f+D}(\hat{K}_*)\leq e_{f+D}(K)$,
\item $e_{f+D}(K)= e_{f+D}(\hat{K}_*)\Rightarrow K\subseteq
\hat{K}_*$.\hfill $\Box$\\
\end{enumerate}}\end{proposition}

{\em Proof}: Intuitively $\hat{K}_*$ is the union of the set of minimum earliness controllable job satisfaction sublanguages. The proof is straightforward. \hfill $\blacksquare$ \\

We call $\hat{K}_*$ the \emph{supremal minimum-earliness
controllable job satisfaction sublanguage of $(G,f+D,h)$ under
$E$ and $\mathcal{E}_T$}, denoted as
$\sup\mathcal{CF}(G,f+D,h,E,\mathcal{E}_T)$. In the remaining of this paper, when we write $\sup\mathcal{CF}(G,f+D,h,E,\mathcal{E}_T)$, we implicitly assume $\textrm{sup}\mathcal{C}(G,f+D,h,E,\mathcal{E}_T)\neq\varnothing$ and thus $\sup\mathcal{CF}(G,f+D,h,E,\mathcal{E}_T)$ exists. The problem of ensuring minimum earliness of job completions by adding proper delays is formalized below.\\

\begin{problem}\label{prob2} If Problem \ref{prob1} has a non-empty solution, i.e., $\textrm{sup}\mathcal{C}(G,f,h,E,\mathcal{E}_T)\neq \varnothing$, then compute a delay function $D^* \in {\bf DC}$ such that
$e_{f+D^*}(\textrm{sup}\mathcal{CF}(G,f+D^*, h,E, \mathcal{E}_T))=\min_{D \in {\bf DC}} e_{f+D}(\textrm{sup}\mathcal{CF}(G, f+D, h,E, \mathcal{E}_T))$.\hfill $\Box$\\ \end{problem}

To solve this problem, we introduce the following concept.\\

\begin{definition}
A non-empty controllable sublanguage $K_1$ of $K$, where $K \subseteq L_m(G)$, with respect to $G$, if it exists, is said to be minimal if for any non-empty controllable sublanguage $K_2 \subseteq K$, $K_2 \subseteq K_1$ implies $K_2=K_1$. \hfill $\square$\\
\end{definition}

Given a finite, non-empty language $K \subseteq L_m(G)$, let ${\bf CL}(K)$ denote the collection of all non-empty minimal controllable sublanguages contained in $K$ with respect to $G$. We need the following definitions and operations. A state {{ } $q$} in a finite state automaton is said to be a controllable (respectively, an uncontrollable) state if all the transitions out of {{ } $q$} are controllable (respectively, uncontrollable). It is said to be a mixed state if there are both controllable and uncontrollable transitions out of state {{ } $q$}. The controllable transitions out of each mixed state are removed and each mixed state is thus transformed into an uncontrollable state. It is easy to see that this preprocessing stage does not affect the computation of ${\bf CL}(K)$. Given a set of languages $\mathcal{L}$, we define a new set of languages $a\mathcal{L}:=\{aL \mid L \in \mathcal{L}\} \subseteq 2^{\Sigma^*}$ for each $a \in \Sigma$. For two sets of languages $\mathcal{L}_1, \mathcal{L}_2$, we define another set of languages $\mathcal{L}_1 \sqcup \mathcal{L}_2=\{L_1 \cup L_2 \mid L_1 \in\mathcal{L}_1, L_2 \in \mathcal{L}_2\}\subseteq 2^{\Sigma^*}$. We propose the following algorithm to compute ${\bf CL}(K)$ when $K$ itself is controllable with respect to $G$. \\
\begin{algorithm}
\caption{Algorithm for Computing ${\bf CL}(K)$}
\label{algorithm2}
\begin{enumerate}
\item \textbf{Input}: A trim tree-structured automaton $G_K \in \phi(\Sigma)$ such that $L_m(G_K)=K$
\item Assign each leave node $q$ of $G_K$ with $\mathcal{L}(q)=\{\{\epsilon\}\}$.
\item For each node $q$ whose children have been assigned a set of languages, do the following until the root node is reached:
  \begin{enumerate}
     \item if $q$ is a controllable state but not a final state, such that {{ } $(q, a_1, q_1), (q, a_2, q_2), \ldots, (q, a_k, q_k)$} is its branches, then $\mathcal{L}(q):=\bigcup_{i \in [1, k]}a_i\mathcal{L}(q_i)$;
     \item if $q$ is both a controllable state and a final state, then $\mathcal{L}(q):=\bigcup_{i \in [1, k]}a_i\mathcal{L}(q_i) \cup \{\{\epsilon\}\}$;
     \item if $q$ is an uncontrollable state, then $\mathcal{L}(q):=\bigsqcup_{i \in [1, k]}a_i\mathcal{L}(q_i)$.
        \end{enumerate}
\item \textbf{Output} $\mathcal{L}(q_0)$, where $q_0$ is the root of $G_K$.
\end{enumerate}
\end{algorithm}

\begin{proposition}
When $K$ is controllable with respect to $G$, ${\bf CL}(K)=\mathcal{L}(q_0)$. \hfill $\Box$ \\
\end{proposition}

{\em Proof}: It is not difficult to {{ } show by structural induction that}, $\mathcal{L}(q)$ is the set of minimal marked behaviors that can be enforced by a supervisor {{ } on $G_K$ when the initial state of $G_K$ is $q$}. Then the set of minimal controllable sublanguages of $K$ will be $\mathcal{L}(q_0)$, i.e., ${\bf CL}(K)=\mathcal{L}(q_0)$.  \hfill $\blacksquare$ \\

With the above technical preparation, a solution to Problem \ref{prob2} is provided in Algorithm~\ref{algo: opt}. It is likely that two different delays lead to the same earliness associated with two different supremal minimum-earliness controllable job satisfaction sublanguages. Algorithm~\ref{algo: opt} indeed computes the set of all such optimal delays. We need the following lemma.\\

\begin{lemma}
\label{lemm: linear}
Let $s$ be any string of $L_m(G)$. For each $j \in [1, n_h]$, {{ } there exist $c_j^i(s, D) \in \mathbb{N}^{|\mathcal{T}|}$ and $c_{j, 0}^i(s, D) \in \mathbb{R}^+ \cup \{0\}$, both of which are functions that depend on $D$, such that} $t_{E_i, f+D}(s)=\max_{j \in [1, n_h]}((c_j^i(s, D))^TD+c_{j, 0}^i(s, D))$, where $D$ is a $|\mathcal{T}|$-tuple vector corresponding to the delay function. \hfill $\Box$ \\
\end{lemma}

{\em Proof}: {{ } This immediately follows from the theory of heaps-of-pieces and the definition of job execution time. In fact, $t_{E_i, f+D}(s)$ is the maximum of the resource time among the $n_h$ different resources and here the $j$-th resource time after the insertion of delay is representable by $(c_j^i(s, D))^TD+c_{j, 0}^i(s, D)$, where the only involved operations are weighted additions of firing duration of each transition (after delay insertion). The reason why both $c_j^i(s, D)$ and $c_{j, 0}^i(s, D)$ depend on $D$ is because the delay function influences the choice of the maximum in the computation of each resource time. } \hfill $\blacksquare$ \\


\begin{algorithm}
\caption{Algorithm for Computing Optimal Delays}
\label{algo: opt}
\begin{enumerate}
\item  \textbf{Input}: $K=\textrm{sup}\mathcal{C}(G,f,h,E,\mathcal{E}_T)$ from Problem 1.
\item  Compute ${\bf CL}(K)$ using Algorithm~\ref{algorithm2}.
\item  For each $L \in {\bf CL}(K)$, solve the following optimization problem:
\[\min_{D\in {\bf DC}}e_{f+D}(L)=\min_{D\in {\bf DC}}\max_{s \in L}e_{f+D}(s)\]
subject to\\
$$\bigwedge_{s \in L} \bigwedge_{i \in I}t_{E_i, f+D}(s) \leq d_i$$  
Let ${\bf DC^*}(L)$ denote the collection of optimal delays for $L$ and $e^*(L)$ denote the minimum earliness that is achievable by $L$ with delays.
\item Let ${\bf OL}:=\{L \in  {\bf CL}(K)\mid \forall L' \in {\bf CL}(K), e^*(L) \leq e^*(L')\}$.
Set ${\bf DC^*}=\bigcup_{L \in {\bf OL}}{\bf DC^*}(L)$.
\item \textbf{Output}: ${\bf DC^*}$.
\end{enumerate}
\end{algorithm}


{{ } According to the monotonicity property of the earliness function, we only need to look at those non-empty minimal controllable sublanguages of $\textrm{sup}\mathcal{C}(G,f,h,E,\mathcal{E}_T)$, i.e., the languages in ${\bf CL}(\textrm{sup}\mathcal{C}(G,f,h,E,\mathcal{E}_T))$, and compute the optimal delays for them, which is shown in Step 3) of Algorithm~\ref{algo: opt}. We have the following theorem.} \\ 

\begin{theorem}\textnormal{Given a plant $(G,f,h)$ and requirements $E$ and $\mathcal{E}_T$, if $\textrm{sup}\mathcal{C}(G,f,h,E,\mathcal{E}_T)\neq\varnothing$, then Algorithm~\ref{algo: opt} solves Problem 2. } \hfill $\Box$ \\
\end{theorem}

{\em Proof}: {{ } Let $D^*$ be a solution of the optimization problem $\min_{D \in {\bf DC}} e_{f+D}(\textrm{sup}\mathcal{CF}(G, f+D, h,E, \mathcal{E}_T))$. According to the definition of the earliness function, there exists a non-empty controllable sublanguage $L \in {\bf CL}(\textrm{sup}\mathcal{CF}(G, f+D^*, h,E, \mathcal{E}_T))$ such that $e_{f+D^*}(L) \leq e_{f+D^*}(\textrm{sup}\mathcal{CF}(G, f+D^*, h,E, \mathcal{E}_T))$. It is not difficult to see that $L \in{\bf CL}(\textrm{sup}\mathcal{C}(G,f,h,E,\mathcal{E}_T))$ and $e^*(L) \leq e_{f+D^*}(L)$. Thus $\min_{L \in {\bf CL}(\textrm{sup}\mathcal{C}(G,f,h,E,\mathcal{E}_T))}e^*(L) \leq e_{f+D^*}(\textrm{sup}\mathcal{CF}(G, f+D^*, h,E, \mathcal{E}_T))$. On the other hand, for any $L \in {\bf CL}(\textrm{sup}\mathcal{C}(G,f,h,E,\mathcal{E}_T))$, let $D^*(L)$ be any solution of the optimization problem $\min_{D\in {\bf DC}}e_{f+D}(L)$ subject to the constraint $\bigwedge_{s \in L} \bigwedge_{i \in I}t_{E_i, f+D}(s) \leq d_i$. Clearly, $e^*(L)=e_{f+D^*(L)}(L) \geq e_{f+D^*}(\textrm{sup}\mathcal{CF}(G, f+D^*, h,E, \mathcal{E}_T))$, since $L \in {\bf CL}(\textrm{sup}\mathcal{C}(G,f,h,E,\mathcal{E}_T))$. Thus $\min_{L \in {\bf CL}(\textrm{sup}\mathcal{C}(G,f,h,E,\mathcal{E}_T))}e^*(L) \geq e_{f+D^*}(\textrm{sup}\mathcal{CF}(G, f+D^*, h,E, \mathcal{E}_T))$.}
\hfill $\blacksquare$  \\

{{ } It is possible to compute the set of all optimal delays by reducing the optimization problem to solving a set of linear programs. The MATLAB toolbox {\bf fminimax} for solving minimax constraint problem can be used for efficiently computing optimal delays with different initializations of $D$.}

For each $D \in {\bf DC^*}$, we could also compute the corresponding supremal minimum-earliness controllable job satisfaction sublanguage $\sup\mathcal{CF}(G,f+D,h,E,\mathcal{E}_T)$ by Proposition~\ref{supre}. We remark that for different delays, the supremal minimum earliness controllable job satisfaction sublanguages could be different. 



\section{Relaxations of Job Deadlines}
If $\textrm{sup}\mathcal{C}(G,f, h,E, \mathcal{E}_T)=\varnothing$, then there is no controllable sublanguage of $L_m(G)$ that satisfies all jobs and the imposed deadlines. If the supremal controllable sublanguage of $L_m(G)\cap E\cap (||_{i\in I}E_i)$ is non-empty, then it means that some job deadlines need to be relaxed in order to have a non-empty controllable job satisfaction sublanguage. Thus it is practically important to know the maximal sets of job deadlines that can be met. Dually, we would like to compute the minimal sets of job deadlines that need to be relaxed. Formally, the following problem is considered:\\
\begin{problem}\label{prob3} Given a plant $(G,f,h)$ and requirements $E$, $\mathcal{E}_T$, determine all the subsets of job requirements $I_1 \subseteq I$ such that\\
\begin{enumerate}
\item $\textrm{sup}\mathcal{C}(G,f, h,E\cap (||_{i\in I_1}E_i), \{(E_i,d_i) \mid i\in I-I_1\})\neq\varnothing$
\medskip
\item $\forall I_2 \subseteq I_1, (\textrm{sup}\mathcal{C}(G,f, h,E\cap (||_{i\in I_2}E_i), \{(E_i,d_i) \mid i\in I-I_2\})\neq\varnothing\Rightarrow I_1=I_2$).\hfill $\Box$ \\
\end{enumerate}
\end{problem}

Note that $\textrm{sup}\mathcal{C}(G,f, h,E\cap (||_{i\in I_1}E_i), \{(E_i,d_i) \mid i\in I-I_1\})$, denoted as $\textrm{sup}\mathcal{CR}(I_1)$ for convenience, is in general an infinite language, since the new job requirements $\mathcal{E'}_T=\{(E_i,d_i) \mid i\in I-I_1\}$ does not satisfy $\Sigma=\bigcup_{i \in I-I_1}\Sigma_i$. {{ } The system is allowed to continuously generate events in $\Sigma -\bigcup_{i \in I-I_1}\Sigma_i$ after all the jobs have been completed. The tree-structured automaton construction method does not work in this case.} We propose the following algorithm, which is quite similar to Algorithm~\ref{algorithm1}, to compute $\textrm{sup}\mathcal{CR}(I_1)$.  \\

\begin{algorithm}
\caption{Computation of $\textrm{sup}\mathcal{CR}(I_1)$}
\label{algo: IV}
\begin{enumerate}
\item \textbf{Input}: A centralized model $(G,f,h)\in\varphi(\Sigma) $, requirements $E\subseteq \Sigma^*$, $\{E_i \mid i \in I_1\}$ and $\mathcal{E'}_T$.
\item Compute the minimal finite automaton $G_L$ for $L_m(G) \cap E \cap \lVert_{i \in I}E_i$. 
\item Construct a finite automaton $A_T$ over $\Sigma$ by constructing a tree-structured automaton for $\Sigma^*$ with the following modifications, such that each string accepted by $A_T$ satisfies the job requirements $I-I_1=\{1, 2, \ldots, k\}$:
\begin{enumerate}
\item Each state $q$ is maintained with a $(k+2)$-tuple $(\hat{\mathcal{M}}_{G,f,h}(s), \upsilon_f(s), t_1(q), t_2(q), \ldots, t_k(q))$, where $s$ is the unique string labeling the path from the root state $q_0$ to state $q$. Initially, $t_1(q_0)=t_2(q_0)=\ldots=t_k(q_0)=0$. 
\item For the immediate successor state $q'$ of $q$ via $\sigma$, let $\hat{\mathcal{M}}_{G,f,h}(s\sigma):=\hat{\mathcal{M}}_{G,f,h}(s)\hat{\mathcal{M}}_{G,f,h}(\sigma)$ and $\upsilon_f(s\sigma):=\textbf{1}_{n_h}^t\hat{\mathcal{M}}_{G,f,h}(s\sigma)\textbf{1}_{n_h}$. If $\sigma \in \Sigma_i$, let $t_i(q'):=\upsilon_f(s\sigma)$, else let $t_i(q'):=t_i(q)$. If $\forall i \in I-I_1, v_f(s\sigma) \geq d_i$, loop at state $q'$ for each $\sigma \in \Sigma-\bigcup_{i \in I-I_1}\Sigma_i$. If {{ } $\exists i \in I-I_1, t_i(q) > d_i$}, the subtree rooted at state $q$ is cut.
\item Repeat step b) for each reached but unexplored state $q$. 
\end{enumerate}
\item Compute $H:=L_m(G_L \lVert A_T)$.
\item \textbf{Output}: The supremal sublanguage $K_*\subseteq H$ that is controllable w.r.t. $G$.
\end{enumerate}
\end{algorithm}
\medskip
{{ } Intuitively, $A_T$ tracks all the strings in $\Sigma^*$ that could satisfy all the job deadlines in $I-I_1$. The following result is immediate.}\\
 
\begin{proposition}
The step 2) of Algorithm~\ref{algo: IV} terminates and the output $K_*$ satisfies $K_*=\textrm{sup}\mathcal{CR}(I_1)$.\hfill $\square$\\
\end{proposition}

{\em Proof}: The proof is straightforward.  \hfill $\blacksquare$ \\

Problem~\ref{prob3} is now solved by examining all the elements of the lattice $(2^I, \cup, \cap, I, \emptyset)$, starting from the bottom element $\emptyset$ to the top element $I$. Once an element, i.e, a subset of $I$, is found to be a solution, there is no need to examine all the greater elements. {{ } In general, there could be different solutions to Problem~\ref{prob3} that are incomparable.}

\section{Simple job-shop example}
\begin{figure}[htb]
\centering
\includegraphics[width=0.2\textwidth]{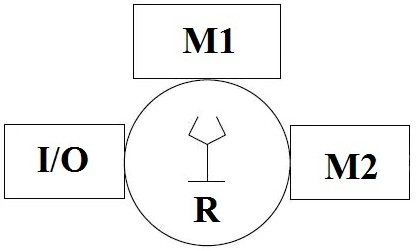}
\caption{Two-Machine Job-Shop Environment}
~\label{job_shop}
\end{figure}

To illustrate above setup, a simplified job shop depicted in Fig.~\ref{job_shop} is used.

\begin{figure}[htb]
\centering
\includegraphics[height=5cm]{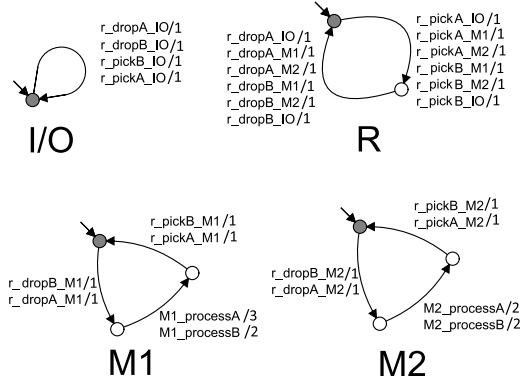}
\caption{Component Model}
~\label{component}
\end{figure}

It consists of one I/O (Input/Output) buffer for feeding
raw materials and dropping out those processed ones. There are two machines (M1 and M2) that process the raw materials, and one robot that performs tasks of loading
and unloading of parts. In this example, we deal with two types of product
($A$ and $B$) and accordingly, based on product type, machines' actions are differentiated.
The timed-weighted model\footnote{In this example, the simple automaton model works because the logic behavior will be further constrained by the job requirement in Fig.3.} for each component is shown in Fig.~\ref{component}.

\begin{figure*}[htb]
\centering
\includegraphics[height=5cm]{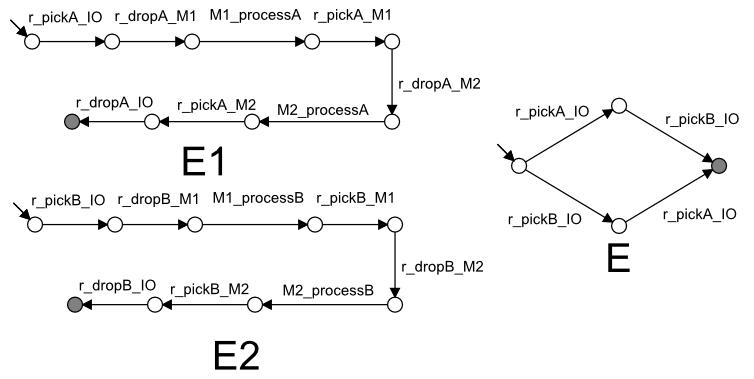}
\caption{Logic Requirement and Job Requirements}
~\label{requirement}
\end{figure*}

\begin{figure*}
\centering
\includegraphics[height=16cm]{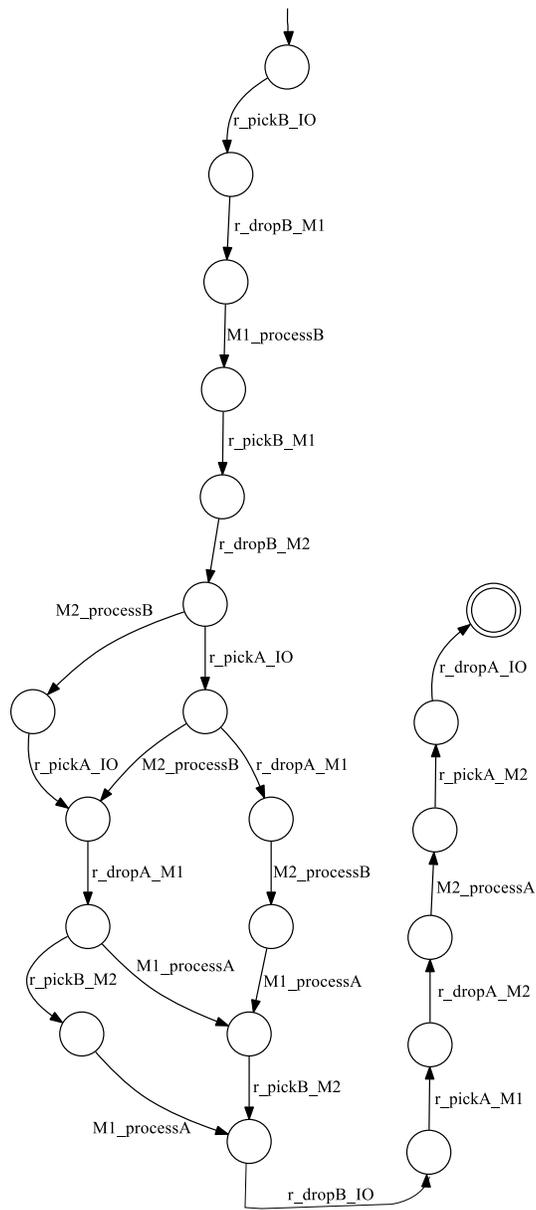}
\caption{Supremal Controllable Job Satisfaction Sublanguage}
~\label{supervisor}
\end{figure*}

We assume that all events of $\Sigma=\Sigma_{M_1} \cup \Sigma_{M_2} \cup \Sigma_R$ are controllable, and the mutual exclusion relation is $h=\Sigma_{M_1}\times \Sigma_{M_1} \cup \Sigma_{M_2}\times \Sigma_{M_2}\cup \Sigma_R\times \Sigma_R $. In this example, to have the complete products, both parts are required to go through $I/O\to M_1\to M_2\to I/O$.
Let $\mathcal{E}_T=\{(E_1,20),(E_2,12)\}$, where $E_1$ and $E_2$ are job requirements for product type $A$ and type $B$ respectively. The logic requirement and job requirements are depicted in Fig.~\ref{requirement}.
Upon applying Algorithm~\ref{algorithm1}, we can compute the supremal controllable job satisfaction sublanguage, which is depicted in Fig.~\ref{supervisor}.
We can easily see that the result of Problem~\ref{prob1} consists of 5 non-empty minimal controllable sublanguages - each one is a singleton. The completion time of $E_1$ and $E_2$ for different strings are as follows: (we order the strings from left to right)

For $i \in [1, 4], t_{J_1, f}(s_i)=16, t_{J_2, f}(s_i)=10$ and for string $s_5, t_{J_1, f}(s_5)=17, t_{J_2, f}(s_5)=11$.

Clearly $e_f(s_i)=6$ for all $i \in [1, 4]$ and $e_f(s_5)=4$, thus $\{s_5\}$  is the supremal minimum-earliness controllable job satisfaction sublanguage. If delay functions are considered, it is easy to see that the  achievable minimum earliness is zero. Indeed, it suffices to check that if we choose $D^*$ such that $D^*(\textrm{r\_dropB\_IO})=1$, $D^*(\textrm{r\_dropA\_IO})=3$, and for all $\sigma \in \Sigma-\{\textrm{r\_dropB\_IO, r\_dropA\_IO}\}, D^*(\sigma)=0$, then $e_{f+D^*}(\sup\mathcal{C}\mathcal{F}(G,f+D^*,h,E,\mathcal{E}_T))=e_{f+D^*}(\{s_5\})=0$. Here, we use events instead of transitions to specify $D$ as there is a clear one to one map between them in this example.

\section{Conclusions}
In this paper we address the problem of timed control for multiple job deadlines.
We have first introduced the concept of supremal controllable job satisfaction sublanguage, and provided algorithms to compute such a supremal sublanguage. Considering that in practical applications large job earliness is usually undesirable, we have introduced the concept of supremal controllable minimum-earliness job satisfaction sublanguage, which ensures minimum job earliness by adding delays in properly chosen transition firings. In case that there does not exist a non-empty controllable job satisfaction sublanguage, we have proposed an algorithm to compute the minimal sets of deadlines that need to be relaxed.

{\em Acknowledgment}:
The support for this work from Ministry of Education Academic Tier 1 Research Grant (No. M4011221 RG84/13) is gratefully acknowledged.

\bibliographystyle{plain}


\end{document}